\documentclass[runningheads]{llncs}
\usepackage[T1]{fontenc}
\usepackage{graphicx}
\usepackage{tcolorbox}
\usepackage{amsmath,amssymb,amsfonts}
\usepackage{algorithmic}
\usepackage{xcolor}
\usepackage{makecell}
\usepackage{booktabs}
\usepackage{multirow}
\usepackage{url}
\usepackage{adjustbox}
\usepackage{caption}
\usepackage{twemojis} % Required for flag emojis like \twemoji{flag: United States}
\usepackage{tabularx}
\usepackage{booktabs}
\usepackage{float}
\usepackage{array}
\PassOptionsToPackage{hyphens}{url}
\usepackage[bookmarks=false]{hyperref}
\newcolumntype{Y}{>{\raggedright\arraybackslash}X}
\newcommand{\model}{STINER}

\begin{document}
\raggedbottom
\title{\model{}: Automated Extraction of Strategic Cyber Threat Intelligence from X}
\titlerunning{\model{}: Automated Extraction of Strategic CTI from X}% If the paper title is too long for the running head, you can set
% an abbreviated paper title here
%
\author{
Yasir ECH\mbox{-}CHAMMAKHY\inst{1} \and
Oussama Azrara\inst{2} \and
Jaafar Chbili\inst{2} \and
Anas Motii\inst{1} 
}
\authorrunning{Y. ECH-CHAMMAKHY et al.}
% First names are abbreviated in the running head.
% If there are more than two authors, 'et al.' is used.
%
\institute{
College of Computing, Mohammed VI Polytechnic University (UM6P), Ben Guerir, Morocco
\and
% Deloitte Morocco Cyber Center, Casablanca, Morocco
% \and
Deloitte Conseil, Paris, France\\
\email{\{Yasir.ECH-CHAMMAKHY,Anas.MOTII\}@um6p.ma, \{oazrara,jchbili\}@deloitte.fr}
}
\maketitle              % typeset the header of the contribution

\begin{abstract}
Strategic Cyber Threat Intelligence (CTI) focuses on high-level insights, such as identifying targeted industries, attributing attacks to specific ransomware groups, and assessing the scale of data loss. Today, X (formerly Twitter) has become the fastest source for this intelligence, often hosting real-time breach announcements days before formal vendor reports. Converting this raw chatter into actionable intelligence requires navigating a complex linguistic landscape. Conventional Named Entity Recognition (NER) models struggle to parse the informal and highly irregular dialect of social media, creating a blind spot for automated defense systems. To address this challenge, we introduce \model{}, a taxonomy and expert-annotated corpus for extracting strategic intelligence from social media streams. We construct a high-quality, expert-annotated dataset of 2,100 real-world alerts and propose a granular taxonomy of eight entity types centered on strategic pivots such as \textit{Threat Actor}, \textit{Sector}, and \textit{Location}. We benchmark nine models across 12 evaluated configurations, spanning general-purpose
and domain-adapted encoders, open-schema extraction, and generative LLMs in both
zero-shot and fine-tuned settings. Domain-adapted encoders such as DarkBERT reach a
strict F1-score of 89.33\%, outperforming both general-purpose baselines and
fine-tuned Large Language Models, which additionally incur substantially higher
inference latency. Leveraging \model{}-DarkBERT, we conduct a European threat landscape analysis for H1 2025. Our results align with official reporting on major targets while highlighting the distinct visibility profile of attacks in \textit{Spain}, and illustrate how social-media-driven extraction can surface early signals of the \textit{SafePay} ransomware campaign prior to its retrospective characterization in vendor threat landscape reports.

\keywords{Cyber Threat Intelligence \and Named Entity Recognition \and Social Media Mining}
\end{abstract}

\section{Introduction}
\label{sec:introduction}

Cyber Threat Intelligence (CTI) is commonly stratified into four layers: Technical, Tactical, Operational, and Strategic~\cite{chismon2015threat}. While the security industry has achieved high maturity in automating the Technical layer, ingesting millions of atomic Indicators of Compromise (IoCs) such as IP addresses and file hashes daily~\cite{barnum2012stix}, a critical gap persists at the Strategic level. As illustrated by the Pyramid of Pain, atomic indicators are trivial for attackers to change and for defenders to detect, whereas strategic insights require contextual reasoning and remain largely resistant to automation~\cite{bianco2013pyramid}.

Strategic Cyber Intelligence (SCI) focuses on the \emph{Who, Why, and How Much} of the threat landscape rather than the technical \emph{What} and \emph{How} \cite{socradar_strategic}. It supports executive decision-making by answering questions such as: \textit{“Is my sector being targeted?”} or \textit{“Which threat actor is active in my region?”} Unlike short-lived technical indicators, SCI informs long-term planning, resource allocation, and organizational risk management.

Traditionally, acquiring this intelligence required manual analysis of long-form vendor reports or access to expensive curated feeds~\cite{arazzi2025nlp,sans2025_cti_survey}. However, the operational tempo of modern threat actors has shifted the primary channel of intelligence dissemination. Today, the earliest and most granular signals of strategic risk, including ransomware victim announcements, initial access auctions, and hacktivist targeting claims, are published first on social media platforms such as X (formerly Twitter), often days before their appearance in formal reports~\cite{cui2025tweezers,arikkat_2025_discerning}.

Leveraging social media for CTI introduces an inherent trade-off. Formal vendor reports take weeks to publish precisely because analysts must rigorously verify facts, filter out bias, and confirm authenticity. In contrast, social media provides extreme velocity but is inherently noisy and unverified. To harness this high-velocity stream without being overwhelmed by noise, automated systems must first isolate and structure the \textit{claims} being made (e.g., a specific ransomware cartel claiming to have breached a targeted sector). While verifying these claims is a downstream task---often requiring cross-referencing with dark web leaks or analyzing social metadata like retweets and sentiment---structuring the unstructured text is the critical bottleneck. 

\begin{figure}[t]
\centering
\includegraphics[width=0.8\textwidth]{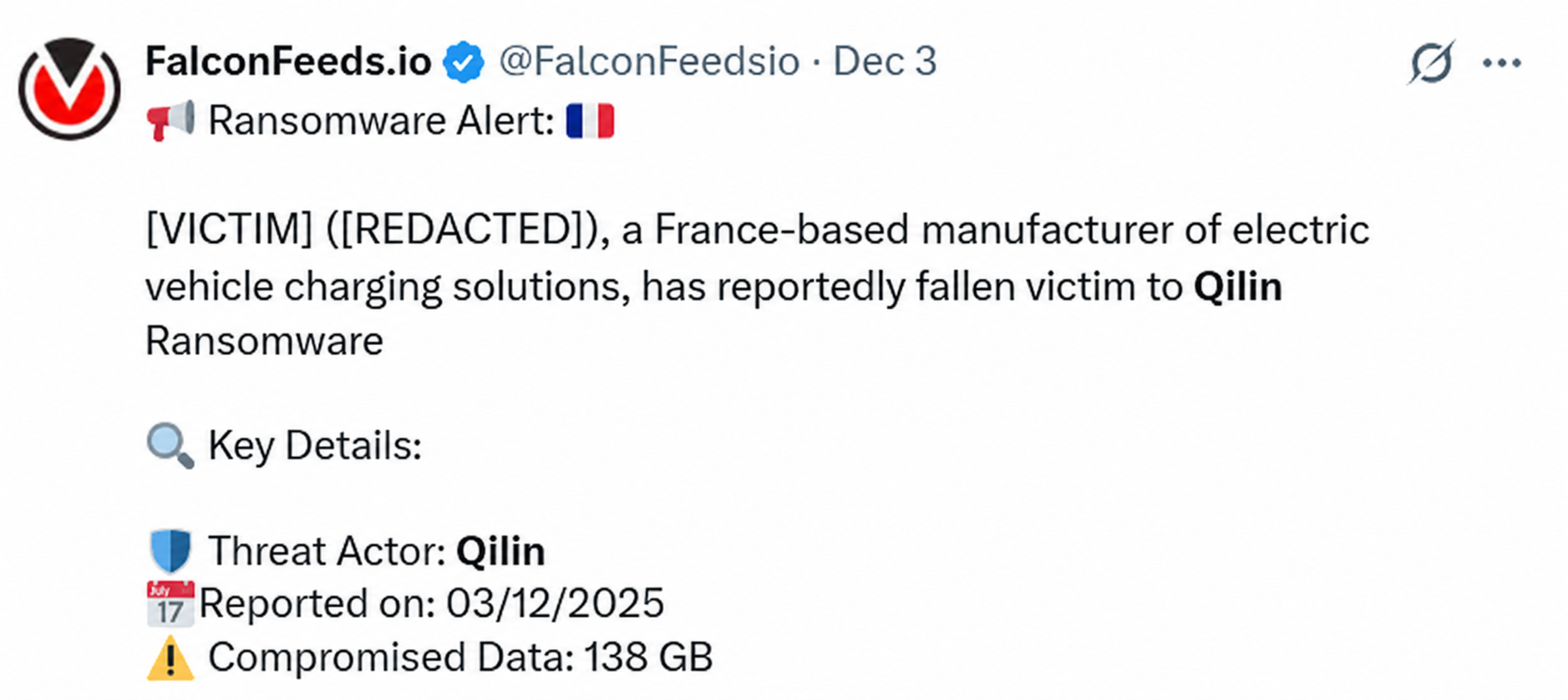}
\caption{A representative CTI alert from Twitter (X). The unstructured text contains dense strategic indicators including Target, Sector, and Actor.}
\label{fig:tweet_example}
\end{figure}

Rather than acting as a simple alerting mechanism, social media has evolved into a consolidated intelligence aggregator \cite{martinsGeneratingQualityThreat2022}. As illustrated in Figure~\ref{fig:tweet_example}, a single stream provides visibility across disparate actors and campaigns, eliminating the need for analysts to manually monitor fragmented sources. This shift enables proactive filtering: defenders can isolate threats targeting specific \textit{Regions} or \textit{Sectors}. To operationalize this capability at scale, robust automated extraction is required.

Named Entity Recognition (NER) is a natural fit for this task, yet existing CTI resources are fundamentally misaligned with the social media domain. Prominent open-source datasets such as DNRTI \cite{wangDNRTILargeScaleDataset2020} and CyNER \cite{alamCyNERPythonLibrary2022} are derived almost exclusively from long-form vendor reports and APT whitepapers. Models trained on such formal prose struggle with the irregular, abbreviated, and hashtag-driven syntax of social media CTI, where attribution is often implicit (e.g., \texttt{\#LockBit}) and targeting is rarely stated explicitly. Prior work on social media CTI has produced annotated corpora, but these target
either binary relevance classification~\cite{behzadan2018corpus,le2019novelty,simran2020deep}
or technical entity extraction focused on affected products, versions, and
vulnerabilities~\cite{dionisio2019cyberthreat}. To the best of our knowledge, no
publicly available, expert-annotated social media corpus targets the
\emph{strategic} layer---victim organization, sector, and quantified breach
impact---which is the level at which organizational risk decisions are made.

To address this gap, we introduce \model{} (Strategic Threat Intelligence Named Entity
Recognition), a resource for extracting strategic CTI from social media. \model{} has
three parts: a taxonomy of eight strategic entity types, an expert-annotated corpus of
2,100 CTI tweets, and a benchmark that reports how well encoder, open-schema, and
generative models perform on it. \model{} does not include a model of its own. It
defines what to extract and provides the labelled data needed to train an extractor,
leaving the choice of architecture open. Throughout the paper, \model{} refers to this
resource, and \model{}-DarkBERT refers to the best-performing extractor we trained on
it, which we use for the analyses in Section~\ref{sec:threat_landscape}.

Our contributions are threefold:
\begin{itemize}
    \item \textbf{A Strategic-CTI NER Dataset for Social Media:} We release a curated
corpus of 2,100 expert-annotated CTI tweets covering eight strategic entity types,
together with the upstream \emph{CTI-relevant} and \emph{CTI-irrelevant}
classification partitions used during dataset construction. Unlike prior social media
CTI corpora, which annotate relevance or technical indicators, \model{} annotates
victim, sector, and impact entities. The dataset, annotation guidelines, and
experimental resources are publicly available at
\url{https://github.com/ChammakhYasir/STINER}.
    
    \item \textbf{Extensive Benchmarking:} We present a rigorous comparison between domain-adapted encoders and generative decoders, demonstrating that specialized models such as DarkBERT achieve superior extraction accuracy (89.33\% strict F1) on noisy social media text while maintaining millisecond-level inference latency.
    
    \item \textbf{European Threat Landscape Analysis:} We leverage \model{} to reconstruct the European strategic threat profile for H1 2025, corroborating official reporting on top-tier targets (Germany, Italy, Spain). Crucially, we demonstrate how structuring social media claims provides actionable lead time, successfully identifying the escalation of the SafePay ransomware campaign prior to its inclusion in formal vendor reports.
\end{itemize}

\section{Related Work}
\label{sec:related_work}

The automation of Cyber Threat Intelligence (CTI) has evolved from simple keyword filtering toward semantic extraction pipelines, driven largely by the availability of annotated corpora that capture domain-specific threat language.

Early CTI NER efforts relied on long-form threat reports. Datasets such as
DNRTI~\cite{wangDNRTILargeScaleDataset2020} and the work of Gao et
al.~\cite{gao2021cyberner} trained BiLSTM-CRF models on vendor whitepapers to extract
malware names and technical indicators. More recent resources, including
CyNER~\cite{alamCyNERPythonLibrary2022} and the harmonized CyberNER
corpus~\cite{echchammakhy2025cyberner}, standardized these efforts under schemas such
as STIX~2.1~\cite{oasis_stix_intro}. However, these datasets share a critical
limitation: they are derived almost exclusively from formal, long-form reports. This
creates a substantial domain mismatch when models are applied to social media CTI,
where language is abbreviated, attribution is implicit (e.g., hashtags), and visual
delimiters such as emojis play a semantic role. Models trained on vendor prose
struggle to generalize to this informal, high-velocity setting, particularly for
strategic entities such as victim sector or breach impact.

A second line of work has targeted social media directly. Behzadan et
al.~\cite{behzadan2018corpus} released an annotated Twitter corpus for cyber threat
relevance classification, and Le et al.~\cite{le2019novelty} framed the same task as
novelty detection against CVE descriptions, avoiding the sampling bias introduced by
negative training samples. Simran et al.~\cite{simran2020deep} extended this line with
deep text-representation methods for tweet-level threat classification. Closest to our
work, Dion\'isio et al.~\cite{dionisio2019cyberthreat} paired a CNN relevance
classifier with a bidirectional NER model over tweets, demonstrating that token-level
extraction from social media CTI is feasible. More recently, Arikkat et
al.~\cite{arikkat2025telegram} constructed a large-scale CTI dataset from curated
Telegram channels using a BERT-based relevance filter, extending social-media CTI
collection beyond Twitter. These efforts establish social media as a viable CTI
source, but their annotation schemas remain oriented toward the technical layer:
entities denote affected products, versions, and vulnerabilities, supporting patch
prioritization rather than strategic risk assessment. \model{} is therefore
complementary rather than competing, annotating who was victimized, in which sector
and region, and at what quantified cost.

Parallel work in NLP has addressed noisy user-generated text through benchmarks such as WNUT17~\cite{derczynski2017wnut} and models like BERTweet~\cite{nguyen_2020_bertweet}, demonstrating that medium-specific pre-training substantially improves robustness to irregular syntax. However, these models lack cybersecurity semantics. For example, a WNUT-trained model may correctly identify “LockBit” as an entity but misclassify it as a generic organization, or fail to distinguish between a company being mentioned and being the victim of a breach. This exposes a persistent gap: existing resources address either social media syntax or cybersecurity semantics, but not both.

Against this backdrop, recent advances in Large Language Models (LLMs) have prompted renewed interest in whether generative architectures can bridge this gap without task-specific training. Their strong performance on a wide range of language understanding tasks has led to growing exploration of LLMs for CTI extraction, often under the assumption that instruction-following alone may obviate the need for specialized NER systems. However, empirical evidence increasingly challenges this assumption. Mezzi et al.~\cite{mezzi2025llm_cti} show that models such as GPT-4 and Claude exhibit inconsistent extraction behavior and hallucinations when applied to CTI. Shafee et al.~\cite{shafee2025llm_osint} report similar findings in OSINT, where LLMs perform well on classification but underperform on token-level extraction. The AthenaBench evaluation~\cite{alam2025athenabench} further highlights reasoning limitations on complex threat scenarios. Beyond accuracy, the latency of autoregressive decoding renders LLMs impractical for real-time processing of high-volume social media streams.

Beyond social media, underground forums and marketplaces constitute a complementary
CTI source with markedly different visibility properties. Prior work has clustered and
ranked security events extracted from hacker forum
discussions~\cite{ech-chammakhy_2025_eventhunter}, showing that actionable signals can
be recovered from adversarial community text. A parallel line of work characterizes the
actors themselves rather than the events, ranking influential participants in
cybercrime forums using graph-based and representation-learning
methods~\cite{hassaneamadou_2024_eurekha,amadou_2026_can,amadou_2024_hchackerrank}. These sources trade the
velocity and breadth of public social media for depth of adversarial context, and we
treat cross-platform integration as future work (Section~\ref{sec:conclusion}).

Taken together, prior work reveals a structural gap in CTI automation: report-centric datasets fail on social media syntax, social-media NLP lacks domain semantics, and LLMs remain both imprecise and operationally inefficient. \model{} addresses this gap by combining a social-media-native dataset with a strategic CTI taxonomy, enabling precise, low-latency extraction that outperforms both generic LLMs and report-trained baselines.

\section{\model{} Dataset Construction}
\label{sec:dataset_construction}

This section describes the construction of the \model{} corpus, including the strategic entity taxonomy, the data collection process, and the expert annotation workflow used to ensure consistent and reliable labeling.

\subsection{Taxonomy Definition}
\label{subsec:taxonomy}

While established frameworks such as MITRE ATT\&CK and STIX 2.1 are global standards for modeling tactical cyber threat intelligence (e.g., capturing specific exploitation vectors, malware families, and atomic indicators), they are fundamentally misaligned with the nature of social media disclosures. Public extortion claims and Initial Access Broker (IAB) auctions rarely detail technical procedures; instead, they focus on economic leverage and business impact. Therefore, rather than forcing high-level social media chatter into a tactical framework, we deliberately designed the \model{} taxonomy to capture \textit{strategic} intelligence. It reconstructs the core narrative of a security incident: \textit{who} was affected, \textit{where} the incident occurred, and \textit{how severe} the impact was.

At the center of the schema are the \texttt{TARGET} (the victim organization) and the corresponding \texttt{SECTOR} (e.g., Healthcare, Energy), which together allow tracking both individual incidents and broader sector-level trends over time. To support geographic and attributional analysis, we additionally extract the victim \texttt{LOCATION} and the threat \texttt{ACTOR} claiming responsibility.

To characterize incident impact, we annotate quantitative and qualitative indicators of breach severity. These include the \texttt{SIZE} of the exfiltrated data (e.g., ``500~GB''), the \texttt{DATA\_TYPE} involved (e.g., ``Source Code'', ``Emails''), and the \texttt{PRICE} associated with ransom demands or data sales. Together, these entities provide the necessary context for downstream impact assessment and economic analysis. The complete definition of all eight entity types is provided in Table~\ref{tab:stiner_schema}.

\begin{table*}[!t]
\centering
\caption{The \model{} entity taxonomy, organized into core strategic pivots (Target, Actor, Sector, Location) and impact-related context (Size, Data Type, Price, Date).}
\label{tab:stiner_schema}
\small
\renewcommand{\arraystretch}{1.3}
\begin{tabularx}{\textwidth}{@{} l >{\raggedright\arraybackslash}X >{\raggedright\arraybackslash}X @{}}
\toprule
\textbf{Entity Label} & \textbf{Definition} & \textbf{Examples} \\ \midrule
\texttt{TARGET} & The organization, company, or institution affected by the incident. & \textit{Wichita State University}, \textit{Tacoma Engineers} \\ \hline
\texttt{ACTOR} & The ransomware group or threat actor claiming responsibility. & \textit{LockBit}, \textit{PLAY}, \textit{IntelBroker} \\ \hline
\texttt{SECTOR} & The industry or market sector in which the victim operates. & \textit{Healthcare}, \textit{Energy}, \textit{Finance} \\ \hline
\texttt{LOCATION} & The country or region associated with the victim. & \textit{USA}, \textit{Italy}, \textit{France} \\ \hline
\texttt{SIZE} & The reported volume of exfiltrated data or number of records. & \textit{10GB}, \textit{1.4TB}, \textit{Full Dump} \\ \hline
\texttt{DATA\_TYPE} & The category of compromised information or assets. & \textit{PII}, \textit{Source Code}, \textit{SQL Dumps} \\ \hline
\texttt{PRICE} & The ransom demand or sale price associated with the incident. & \textit{\$1500}, \textit{10 BTC}, \textit{500K USD} \\ \hline
\texttt{DATE} & Explicit dates related to disclosure or incident timing. & \textit{2025-02-14}, \textit{October 2024} \\ \bottomrule
\end{tabularx}
\end{table*}

\begin{figure}[t]
\centering
\includegraphics[width=\textwidth]{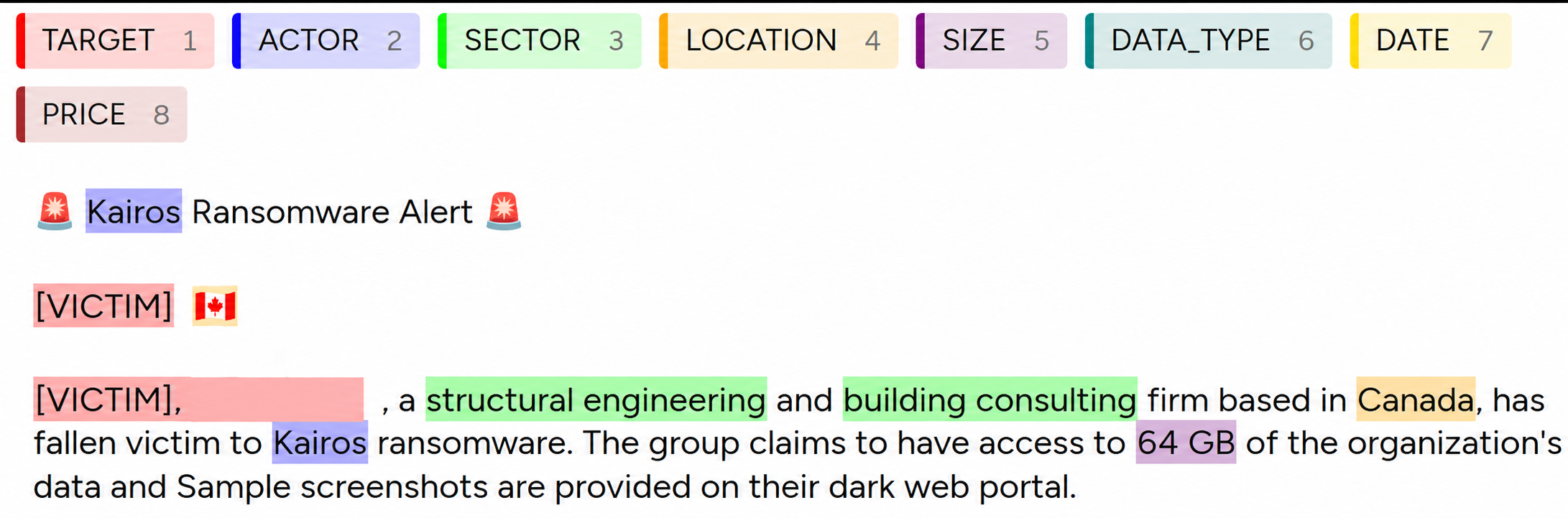}
\caption{Snapshot of the annotation interface used during dataset construction.}
\label{fig:annotation_interface}
\end{figure}

\subsection{Data Collection and Selection}

We collected approximately 66,500 tweets using the Apify~\cite{apify} platform and filtered them for
CTI relevance using a lightweight BERT-based classifier. From this subset, 2,100 tweets
were selected for expert annotation. Selection prioritized source diversity and temporal
coverage, with proportional sampling from independent security researchers and reduced
weighting of high-volume news aggregators. Tweets were sampled uniformly from 2022 to
2025 to avoid temporal bias and reflect evolving reporting practices.

\subsection{Annotation Methodology}

All tweets were manually annotated by two cybersecurity researchers using the Label Studio interface~\cite{labelstudio} (Figure~\ref{fig:annotation_interface}). To assess annotation consistency, 10\% of the dataset (210 tweets) was independently labeled by both annotators, yielding an inter-annotator agreement of Cohen’s~\cite{mchugh2012kappa} $\kappa = 0.84$,
computed at the span level using exact match criteria, requiring exact agreement on both entity boundaries
and labels. Remaining disagreements were resolved through adjudication with a senior reviewer to establish consistent guidelines.

A recurring source of disagreement involved organizations whose official names include geographic descriptors (e.g., ``South Korean Ministry of Environment''). To address this, we defined an \emph{Identity Rule}: when a location is part of an organization’s official name, it is included within the \texttt{TARGET} span; when the location is syntactically separate (e.g., ``based in South Korea''), it is annotated as a distinct \texttt{LOCATION} entity. This rule ensures consistent treatment of named entities across the corpus.

\begin{table}[h]
\centering
\caption{Distribution of entity annotations in the \model{} dataset.}
\label{tab:entity_stats}
\small
\renewcommand{\arraystretch}{1.2}
\begin{tabular}{@{}lrr@{}}
\toprule
\textbf{Entity Type} & \textbf{Count} & \textbf{Frequency (\%)} \\ \midrule
\texttt{LOCATION}    & 4,728          & 27.36 \\
\texttt{TARGET}      & 3,743          & 21.66 \\
\texttt{DATA\_TYPE}  & 3,419          & 19.79 \\
\texttt{ACTOR}       & 2,430          & 14.06 \\
\texttt{SECTOR}      & 1,244          & 7.20  \\
\texttt{SIZE}        & 978            & 5.66  \\
\texttt{PRICE}       & 561            & 3.25  \\
\texttt{DATE}        & 178            & 1.03  \\ \midrule
\textbf{Total}       & \textbf{17,281} & \textbf{100.0} \\ \bottomrule
\end{tabular}
\end{table}

\subsection{Dataset Statistics and Characteristics}
\label{subsec:dataset_stats}

The final \model{} corpus contains 2,100 annotated tweets with a total of 85,066 tokens. Each tweet includes multiple strategic entities, with an average of 8.23 entities per sample. This density is substantially higher than general-purpose social media benchmarks such as WNUT-17~\cite{derczynski2017wnut}, which averages fewer than one entity per tweet.

The dataset vocabulary consists of 7,870 unique tokens, with a Type--Token Ratio of 0.093. This reflects the specialized and repetitive terminology common in cyber threat reporting, where technical terms recur frequently across incidents. As a result, the dataset emphasizes domain-specific concepts rather than conversational language.

As shown in Table~\ref{tab:entity_stats}, most annotations correspond to contextual entities such as \texttt{LOCATION}, \texttt{TARGET}, and \texttt{ACTOR}, which form the foundation of attribution and situational awareness. Impact-related entities such as \texttt{DATA\_TYPE} and \texttt{SIZE} are also well represented, enabling quantitative analysis of breach severity. In contrast, \texttt{PRICE} remains relatively infrequent, reflecting the fact that ransom demands are often negotiated privately and not consistently disclosed in public reports.

% --- SECTION 4: MODEL TRAINING ---
\section{Model Benchmarking}
\label{sec:model_benchmarking}
To identify the optimal architecture for real-time strategic intelligence extraction, we conducted a rigorous comparative study of state-of-the-art Natural Language Processing (NLP) models. This section details our experimental framework, contrasting the efficacy of specialized discriminative encoders against generative approaches under strict operational constraints of precision and latency.

\subsection{Experimental Setup}
\label{subsec:experimental_setup}

We evaluate \model{} using a temporal evaluation protocol designed to reflect real-world deployment conditions. The dataset is split chronologically into training (70\%), validation (15\%), and test (15\%) partitions. Concretely, the training set covers tweets from 2022 to mid-2024, while the test set consists exclusively of tweets from late 2024 to 2025. This ensures that models are evaluated on future, unseen threat campaigns rather than memorizing events from the same time period. Because threat actors, campaign names, and victim organizations evolve rapidly, the 2025 test partition contains threat groups and victim organizations not represented in the training data. This design provides stronger evidence that the framework generalizes based on linguistic context and cybersecurity semantics, rather than merely matching a static dictionary of known actors — directly addressing the risk of evaluation contamination in datasets spanning multiple years.

All models operate directly on raw tweet text, preserving URLs, hashtags, emojis, and user handles, which often convey attributional and campaign-level information in social media CTI.

We benchmark four model families:
\begin{enumerate}
    \item \textbf{General-Purpose Encoders:} BERT-Base~\cite{devlin2019bert} and Twitter-RoBERTa~\cite{barbieri2020tweeteval}.
    \item \textbf{Domain-Specific Encoders:} SecBERT~\cite{secbert_jackaduma_2022}, DarkBERT~\cite{jinDarkBERTLanguageModel2023}, and \allowbreak CySec-BERT~\cite{bayer_2022_cysecbert}.
    \item \textbf{Open-Schema Encoder:} GLiNER-Large-v2.1~\cite{zaratiana_2024_gliner}, evaluated in a zero-shot setting using natural-language label descriptions.
    \item \textbf{Generative LLMs:} Llama-3-8B~\cite{grattafiori_2024_llama}, Gemma-2-9B~\cite{team_2024_gemma}, and Qwen-2.5-14B~\cite{qwen2025qwen25}, evaluated both in a zero-shot configuration (instruction prompting) and after supervised fine-tuning with QLoRA~\cite{dettmers2023qlora}.
\end{enumerate}

Only a subset of these models is trained on \model{}. The five discriminative encoders
(BERT-Base, Twitter-RoBERTa, SecBERT, CySecBERT, DarkBERT) are fine-tuned on the
\model{} training split, as are the three generative models in their QLoRA
configuration. The remaining entries---GLiNER and the three instruction-tuned LLMs in
their zero-shot configuration---never see \model{} training data and are evaluated
directly on the test split using label descriptions or prompting alone. All twelve
entries are evaluated on the same held-out test split, so trained and untrained
configurations are directly comparable.

Encoder-based models are trained using a standard linear token-classification head. We additionally evaluate Conditional Random Field (CRF) decoding as a targeted ablation study (Section~\ref{subsec:crf_ablation}). Generative models employ constrained JSON decoding to enforce a fixed output schema.

Extraction quality is measured using span-level Micro-F1 with a strict matching criterion, requiring exact agreement between predicted and gold entity spans. Details regarding batching strategy, hardware configuration, and latency measurement are reported in Appendix~\ref{app:evaluation_protocol}.

\subsection{Performance Analysis}
\label{subsec:perf_analysis}

Table~\ref{tab:main_results} presents the comparative evaluation of nine models across
12 configurations on the held-out test set. Our analysis reveals critical trade-offs
between domain adaptation, model scale, and operational latency.

\subsubsection{Social Media Syntax vs. Domain Vocabulary}
DarkBERT (89.33\%) and Twitter-RoBERTa (89.23\%) achieve nearly identical performance, indicating that cybersecurity-specific vocabulary alone is not the dominant factor for extraction quality in social media CTI. Instead, robustness to platform-specific syntax—such as hashtags, user handles, abbreviations, and emojis—plays an equally important role.

This effect is highlighted by the weaker performance of SecBERT (Strict F1: 71.69\%). Although SecBERT encodes rich cybersecurity terminology (e.g., CVEs and malware families), its pre-training on formal APT reports limits its ability to generalize to the abbreviated and irregular grammar of tweets. Twitter-RoBERTa, despite lacking explicit security-domain pre-training, benefits from its native exposure to social media language and therefore performs competitively. DarkBERT combines both informal syntax exposure and security-domain semantics, resulting in the strongest overall performance and demonstrating the importance of aligning both the medium and the domain.

\begin{table*}[t]
\centering
\caption{Comprehensive Benchmarking Results. We compare specialized Encoders against Generative LLMs in both Zero-Shot (*) and Fine-Tuned settings. \textbf{Strict F1} requires exact span boundaries. \textbf{Latency} denotes inference time per sample (one tweet).}
\label{tab:main_results}
\small
\begin{tabular}{l c c c c}
\toprule
\textbf{Model Architecture} & \textbf{Strict F1} & \textbf{Precision} & \textbf{Recall} & \textbf{Latency (ms)} \\
\midrule
\multicolumn{5}{l}{\textit{Discriminative Encoders (Fine-Tuned)}} \\
\textbf{DarkBERT} \cite{jinDarkBERTLanguageModel2023} & \textbf{89.33} & 87.76 & \textbf{90.95} & 0.71 \\
Twitter-RoBERTa \cite{barbieri2020tweeteval} & 89.23 & 87.66 & 90.85 & 0.71 \\
CySecBERT \cite{bayer_2022_cysecbert} & 87.02 & 84.20 & 90.04 & 0.71 \\
BERT-Base (Baseline) \cite{devlin2019bert} & 85.01 & 81.85 & 88.42 & 0.71 \\
SecBERT \cite{secbert_jackaduma_2022} & 71.69 & 67.95 & 75.86 & \textbf{0.37} \\
\midrule
\multicolumn{5}{l}{\textit{Open-Schema Models}} \\
GLiNER-Large-v2.1* \cite{zaratiana_2024_gliner} & 71.31 & 74.34 & 68.53 & 35.55 \\
\midrule
\multicolumn{5}{l}{\textit{Generative LLMs (Zero-Shot vs. Fine-Tuned)}} \\
Gemma-2-9B-IT* \cite{team_2024_gemma} & 76.35 & 86.27 & 68.48 & 331.54 \\
Llama-3-8B (Fine-Tuned) \cite{grattafiori_2024_llama} & 74.55 & 78.65 & 70.86 & 1965.94 \\
Gemma-2-9B (Fine-Tuned) \cite{team_2024_gemma} & 72.59 & 82.09 & 65.07 & 3937.96 \\
Qwen-2.5-14B (Fine-Tuned) \cite{qwen2025qwen25} & 66.70 & 82.33 & 56.06 & 6473.18 \\
Qwen-2.5-14B-Instruct* \cite{qwen2025qwen25} & 59.74 & 86.13 & 45.73 & 171.77 \\
Llama-3-8B-Instruct* \cite{grattafiori_2024_llama} & 27.78 & \textbf{89.56} & 16.43 & 110.72 \\
\bottomrule
\end{tabular}
\end{table*}

\subsubsection{The Recall Crisis in Generative AI}
While zero-shot LLMs achieve high precision (86--89\%), they exhibit 
severe recall degradation (16--45\%). This pattern indicates a clear 
recall limitation: generative models reliably extract explicit, 
well-formed entities (e.g., ``LockBit'') but frequently miss 
abbreviated or irregular expressions common in social media CTI 
(e.g., ``\#lockbit3'', ``500GB dump'').

Fine-tuning with QLoRA substantially mitigates this issue. Most 
strikingly, Llama-3 improves from 27.78\% to 74.55\% strict F1 --- 
a 47-point gain that directly quantifies the contribution of 
domain-specific supervised data for social media CTI extraction. 
This demonstrates that instruction prompting alone is insufficient; 
models must be trained directly on the linguistic conventions present 
in real-world social media disclosures to achieve operationally 
usable recall. Critically, even after fine-tuning, the best 
generative model (Llama-3 at 74.55\%) still falls 15 points below 
DarkBERT (89.33\%), while incurring a latency penalty of over 
2{,}700$\times$, making encoder-based architectures the clear 
choice for operational deployment.
\subsubsection{Operational Latency Considerations}
Latency becomes relevant when strategic CTI extraction is embedded within continuous monitoring
or early-warning pipelines. In such settings, encoder-based models process samples in under
1\,ms under batched inference, enabling near-real-time analysis of high-volume social media
streams. In contrast, fine-tuned LLMs incur per-sample latencies of several seconds, restricting
their use to offline or low-throughput analysis. 

As a result, the \model{} dataset is directly compatible with systems that rely on real-time
NER for security event detection and prioritization, such as streaming OSINT pipelines and
early-warning frameworks (e.g., Tweezers~\cite{cui2025tweezers}). In these architectures,
low-latency extraction is a prerequisite, even when the downstream intelligence task is
strategic rather than reactive.

\subsubsection{GLiNER: A Cold-Start Compromise}
GLiNER occupies a distinct operational niche. Although its zero-shot performance (71.3\% F1)
lags behind fine-tuned encoders, it requires no labeled data. This makes it well-suited for
day-zero scenarios, such as emerging threat actors or novel campaign types, where annotated
examples are unavailable. GLiNER offers an order-of-magnitude speed advantage over LLMs while
providing sufficient accuracy for initial triage, serving as a temporary solution until a
specialized encoder can be trained.

\subsection{Ablation Study: Impact of Structured Decoding}
\label{subsec:crf_ablation}

\begin{table}[H]
\centering
\caption{Impact of CRF decoding on encoder-based NER models in terms of effectiveness (Strict Span-Level Micro-F1) and efficiency (latency per sample) on the \model{} test set.}
\label{tab:crf_ablation}
\begin{tabular}{lccccc}
\toprule
\textbf{Model} & \textbf{Linear F1} & \textbf{+CRF F1} & $\Delta$ F1 & \textbf{Latency (ms)} & \textbf{Slowdown} \\
\midrule
BERT-Base        & 0.850 & 0.831 & $-0.019$ & $0.71 \rightarrow 4.84$ & 6.8$\times$ \\
Twitter-RoBERTa  & 0.892 & 0.838 & $-0.054$ & $0.71 \rightarrow 5.05$ & 7.1$\times$ \\
SecBERT          & 0.717 & 0.769 & \textbf{+0.052} & $0.37 \rightarrow 3.32$ & 9.0$\times$ \\
CySecBERT        & 0.870 & 0.849 & $-0.021$ & $0.71 \rightarrow 4.85$ & 6.8$\times$ \\
DarkBERT         & \textbf{0.893} & 0.841 & $-0.052$ & $0.71 \rightarrow 5.06$ & 7.1$\times$ \\
\bottomrule
\end{tabular}
\end{table}

Structured decoding layers such as Conditional Random Fields (CRFs) are widely adopted in
cybersecurity NER pipelines, particularly in report-centric settings, where they have been
shown to improve label consistency by enforcing valid BIO transition constraints
(e.g., \cite{chen2023bertcrf_cti}, \cite{mouiche2025entity_relation_cti}, \cite{sun2020hybrid_crf}).
Motivated by this precedent, we evaluate whether CRF decoding provides similar benefits for
social-media-native CTI extraction.

We compare encoder-based models equipped with a standard linear token-classification head,
which predicts labels independently, against an alternative CRF decoding layer that enforces
structured label transitions (e.g., preventing \texttt{I-ACTOR} from following
\texttt{B-MALWARE}). Results are reported in Table~\ref{tab:crf_ablation}.

The findings reveal a clear accuracy–efficiency trade-off. While CRF decoding improves the
performance of SecBERT (+5.2 F1), likely compensating for weaker contextual representations in
informal text, it consistently degrades performance for stronger encoders. In particular,
DarkBERT and Twitter-RoBERTa experience F1 drops exceeding 5 points when constrained by CRF
decoding.

This behavior reflects a mismatch between CRF assumptions and the structure of social media
CTI. Tweets frequently contain fragmented entities, unconventional boundaries, and
non-canonical grammar that violate strict BIO transition patterns. Modern transformer encoders
implicitly model these irregularities through self-attention, whereas CRF decoding imposes
rigid Markovian constraints that can over-correct semantically valid but syntactically noisy
predictions.

Beyond accuracy, CRF decoding introduces substantial computational overhead. The
non-parallelizable Viterbi algorithm~\cite{xie_viterbi_algorithm} increases inference latency by approximately 7$\times$,
shifting performance from sub-millisecond batched inference to several milliseconds per
sample. As a result, we adopt a linear decoding head in the final \model{} pipeline, balancing
robust extraction accuracy with operational efficiency.

\subsection{Fine-Grained Entity Analysis}
\label{subsec:entity_analysis}

\begin{figure}[t]
\centering
\includegraphics[width=\columnwidth]{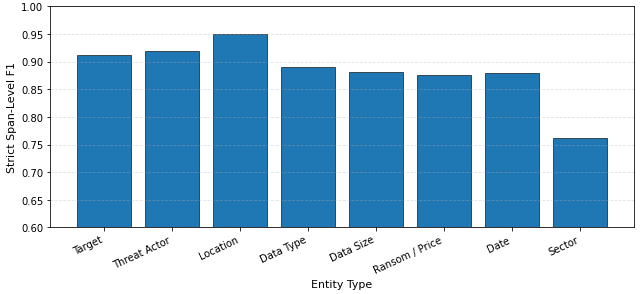}
\caption{Per-entity strict span-level F1 scores on the \model{} test set for the best-performing model (DarkBERT with a linear decoding head).}
\label{fig:per_entity_f1}
\end{figure}

To analyze strengths and failure modes at the entity level, we report per-class performance for
the best-performing configuration, DarkBERT with a linear decoding head.
Figure~\ref{fig:per_entity_f1} shows strict F1 scores across the eight strategic entity types.

The model achieves very strong performance on the core attribution entities:
\texttt{LOCATION} (95.0\%), \texttt{ACTOR} (91.9\%), and \texttt{TARGET} (91.3\%).
These entities are typically expressed using distinctive surface cues in social media CTI,
including flag emojis for countries, standardized ransomware group names, and explicit organization
mentions. Such cues provide clear signals that are effectively captured by transformer-based
attention mechanisms.

In contrast, \texttt{SECTOR} exhibits the lowest performance (76.2\%).
Inspection of prediction errors shows that this is primarily due to contextual ambiguity rather
than extraction failure. In many tweets, the same term may appear either as part of an organization
name (\texttt{TARGET}) or as a description of the victim’s business domain (\texttt{SECTOR}),
depending on context. Resolving this ambiguity often requires knowledge about the organization
itself, which is not always available from the local sentence alone.

Crucially for downstream risk assessment, impact-related entities are extracted with high
reliability. \texttt{DATA\_TYPE} reaches 89.1\% strict F1 and \texttt{SIZE} achieves 88.0\%,
indicating that the model consistently captures both the nature and scale of exfiltrated data.
This enables automated estimation of incident severity based on data sensitivity and volume, a
central requirement for strategic CTI analysis.

\section{European Threat Landscape Analysis (H1 2025)}
\label{sec:threat_landscape}

Given the geopolitical significance of the region and the enforcement of the NIS2 Directive~\cite{eu2026_nis2}, we focused our threat landscape analysis on the European theater. By filtering \model{}-DarkBERT extractions for entities located within the EU-27 and the UK, we reconstructed the strategic risk profile facing European critical infrastructure. Crucially, to assess the validity of our real-time social media extraction, we benchmarked our findings against the authoritative \textit{ENISA Threat Landscape 2025} report ~\cite{enisa2025threat}. This comparison demonstrates that \model{} not only corroborates official retrospective data but often serves as a leading indicator for emerging campaigns.

\subsection{Geographic Targeting Patterns}
\begin{figure}[t]
\centering
\includegraphics[width=0.8\columnwidth]{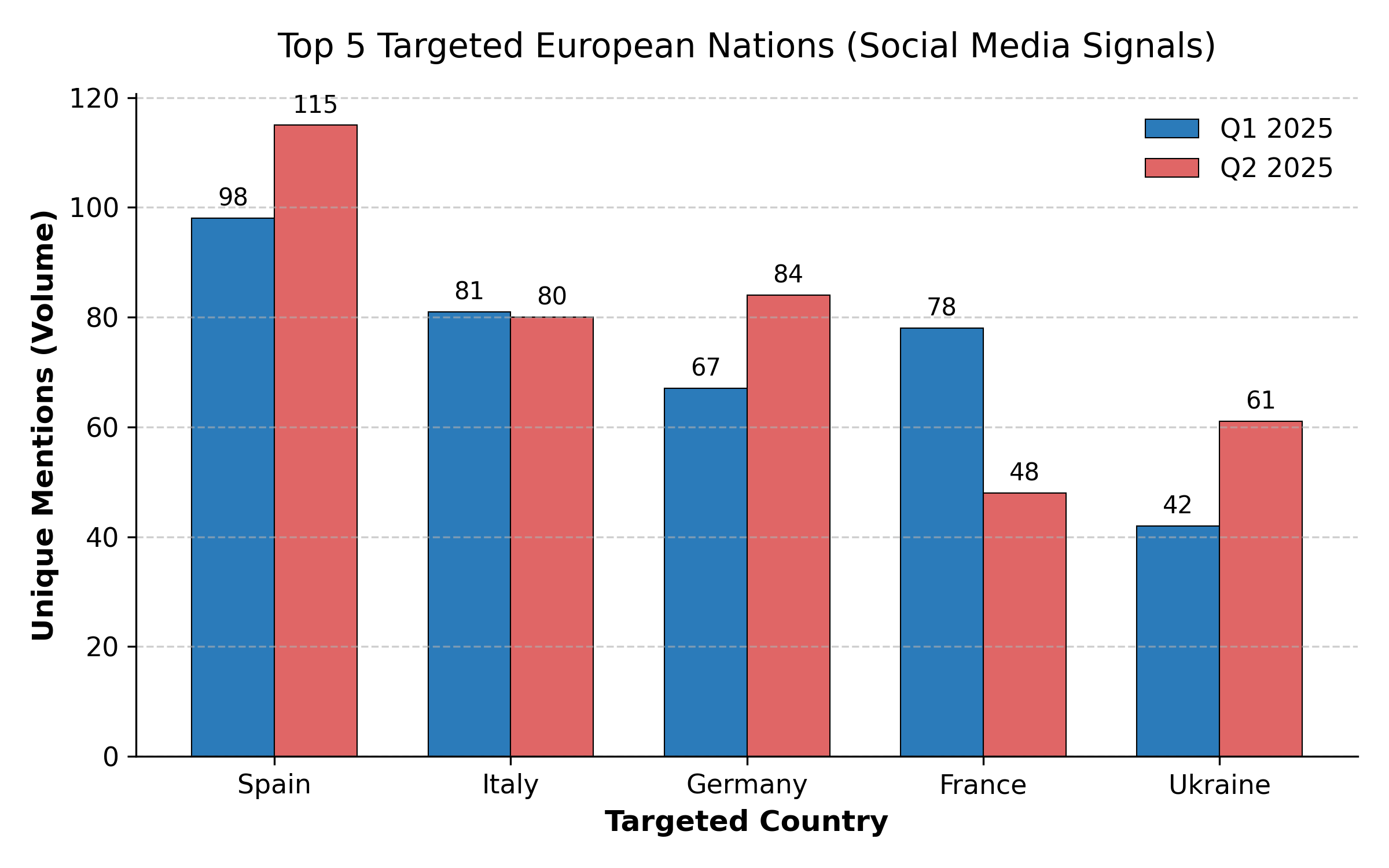}
\caption{Publicly claimed targeting of European nations across H1 2025, comparing Q1 and Q2 trends.}
\label{fig:geo_targeting}
\end{figure}
Figure~\ref{fig:geo_targeting} illustrates the distribution of victim organizations across Europe. Spain emerged as the primary target in our dataset, followed closely by Italy and Germany.

While this finding aligns with official reports in terms of the countries most affected, the ranking differs. The \textit{ENISA Threat Landscape 2025} identifies Germany as the most affected country based on confirmed incidents, whereas Spain appears as the most frequently targeted country in the \model{} dataset derived from social media disclosures. This difference reflects the nature of the underlying data sources: ENISA captures validated incidents, while \model{}-DarkBERT. measures the public visibility of attacks through claims and extortion-related posts. As a result, Spain’s prominence indicates higher disclosure activity rather than a higher absolute incident count.

To validate this hypothesis, we performed a granular extraction of the threat landscape specific to Spain. \model{}-DarkBERT successfully identified Akira and Qilin as the two dominant adversaries targeting the region in H1 2025. This finding is independently corroborated by the \textit{SOCRadar Spain Threat Landscape Report 2025}, which ranks these exact groups as the top ransomware threats \cite{socradar2025spain}. This alignment confirms that \model{}-DarkBERT. not only captures broad volume trends but accurately reconstructs the specific attribution profile of regional cyber conflicts without manual intervention.

\subsection{Temporal Analysis: The SafePay Case Study}
\begin{figure}[t]
\centering
\includegraphics[width=\columnwidth]{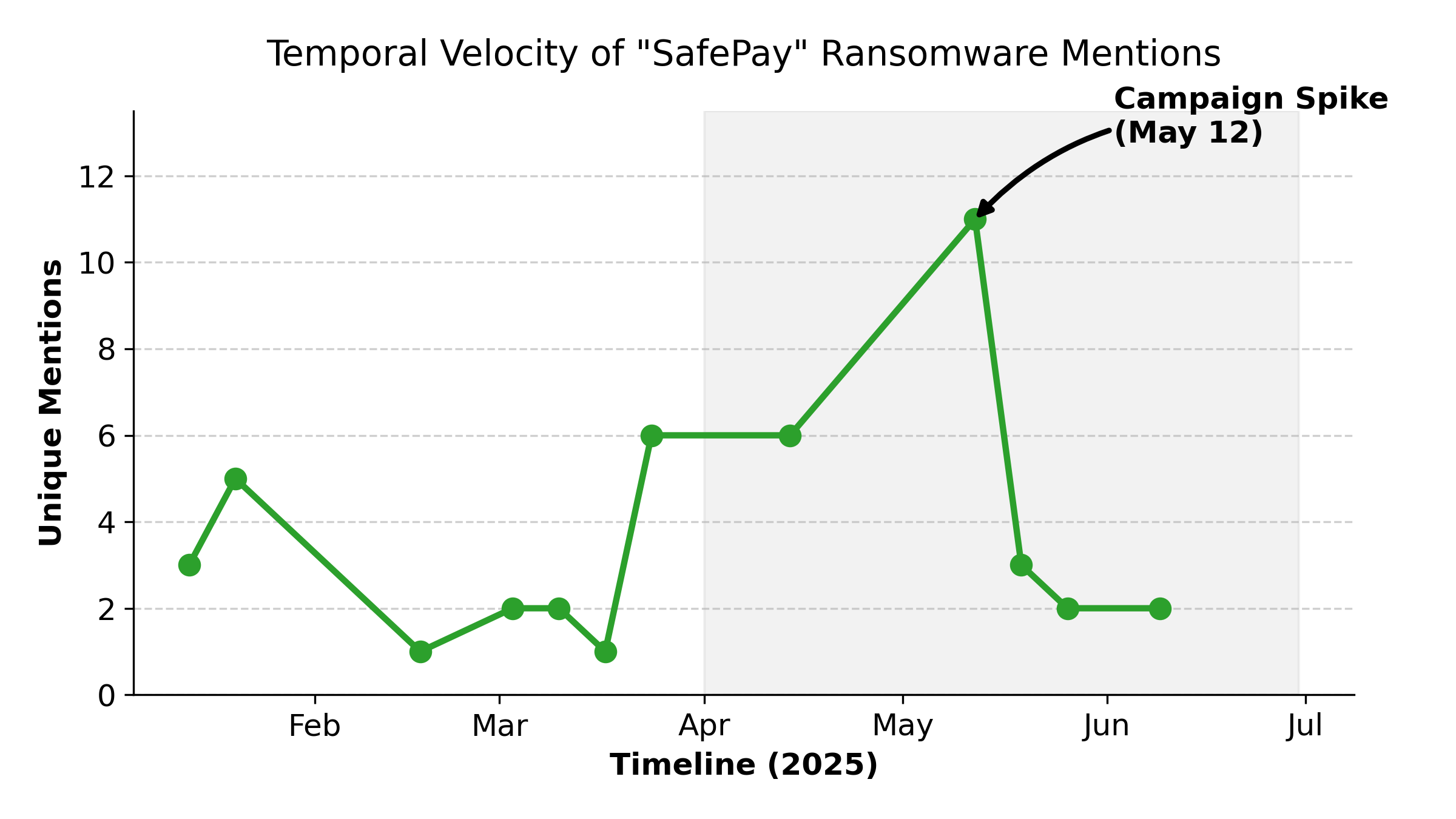}
\caption{Weekly mention velocity for the SafePay ransomware group. 
\model{}-DarkBERT. detects a distinct campaign spike in mid-May 2025, 
approximately 6--7 weeks before SafePay's consolidation as a 
rising threat in the ENISA Q2 2025 reporting period.}
\label{fig:safepay_trend}
\end{figure}

A core motivation for social-media-driven CTI is temporal resolution.
Authoritative threat landscape reports, such as the ENISA Threat 
Landscape 2025~\cite{enisa2025threat}, necessarily operate at a 
quarterly or semi-annual aggregation level, reflecting confirmed and 
curated incidents. In contrast, \model{}-DarkBERT. enables week-level visibility 
into campaign dynamics as they unfold in public discourse.

We select SafePay as a case study because it is explicitly identified 
by ENISA as an emerging ransomware strain that rose to prominence in 
Q2 2025, ranking as the second most deployed strain against EU public 
administration (33.3\% of ransomware claims in that sector)~\cite{enisa2025threat}.

Figure~\ref{fig:safepay_trend} shows that \model{}-DarkBERT detects a sharp 
inflection in SafePay-related activity during the week of May 12--18, 
2025, marked by a sudden increase in public claims and victim 
disclosures on social media. This spike corresponds to the early 
escalation phase of the campaign, when operators actively publicize 
breaches to amplify extortion pressure and visibility. Crucially, 
this detection precedes ENISA's Q2 2025 consolidation of SafePay as 
a rising threat by approximately 6--7 weeks --- a gap that directly 
quantifies the early-warning advantage of social-media-driven CTI 
over institutional reporting cycles. This trajectory is independently 
corroborated by ENISA's quarterly ransomware claims data, which 
documents SafePay's surge as one of the most active strains in the 
EU over the same period~\cite{enisa2025threat}.

This temporal gap highlights a structural complementarity between 
social-media-driven intelligence and institutional threat reporting. 
By detecting inflection points in disclosure velocity, \model{}-DarkBERT. 
enables earlier defensive actions, such as placing the actor under 
heightened monitoring, issuing sector-specific warnings, or 
prioritizing threat hunting for associated infrastructure. In this 
role, \model{}-DarkBERT functions as a leading indicator that anticipates 
trends later confirmed by authoritative threat landscape assessments.

\subsection{Quantifying the Intelligence Gap}

Beyond attribution and timing, a persistent challenge in strategic cyber threat intelligence is quantifying impact. Official bodies, including ENISA, explicitly acknowledge the difficulty of assessing economic and operational damage due to under-reporting and inconsistent disclosure of breach severity \cite{enisa2025threat}.

By aggregating the \texttt{SIZE} entities extracted by \model{}, we reconstructed the map of compromised data volume for H1 2025. While incident counts often highlight public administration due to frequent service disruptions, our volumetric analysis (Table~\ref{tab:sector_impact}) reveals a different reality regarding data loss.

The Manufacturing sector emerges as the primary victim of data exfiltration, with over 146 TB of data leaked, confirming the heavy toll of industrial espionage and extortion on this vertical. Notably, the Government sector also suffered massive data exposure ($>$104 TB), indicating that beyond website disruptions, state entities face significant data confidentiality breaches. This contrast highlights a critical intelligence gap: measuring risk solely by "number of incidents" obscures the true strategic cost. \model{} provides the necessary quantitative dimension to prioritize defenses based on the magnitude of potential data loss.

\begin{table}[h]
\centering
\caption{Sectoral Vulnerability Analysis: Total Leaked Data Volume (H1 2025).}
\label{tab:sector_impact}
\small
\begin{tabular}{lr}
\toprule
\textbf{Sector} & \textbf{Leaked Volume (GB)} \\
\midrule
Manufacturing & 146,018 \\
Government & 104,445 \\
Military & 82,510 \\
Architecture \& Planning & 71,680 \\
Healthcare & 67,365 \\
\bottomrule
\end{tabular}
\end{table}

\subsection{Strategic Threat Profiling: The Medusa Cartel (Q1 2025)}
\label{subsec:medusa_profile}

\begin{figure}[H]
\centering
\includegraphics[width=0.9\linewidth]{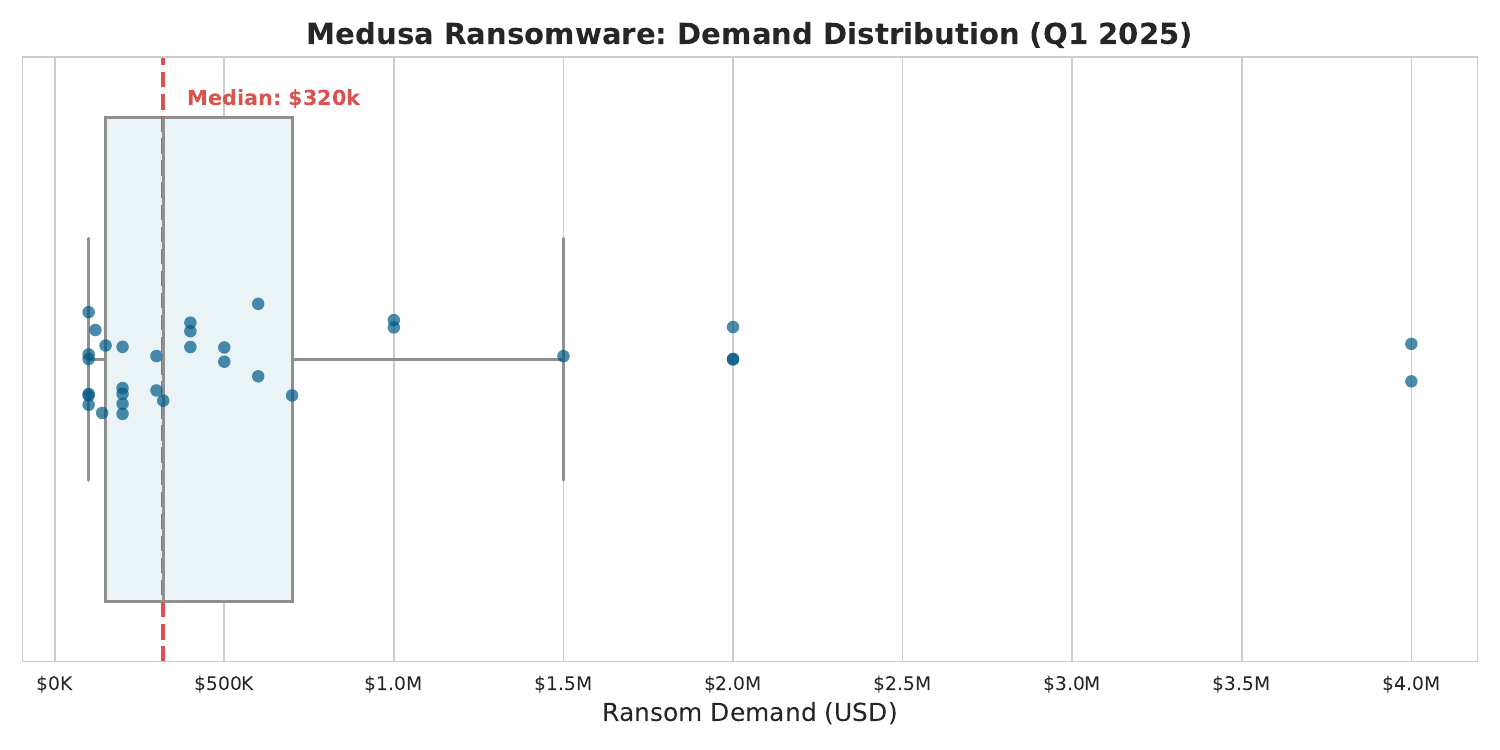}
\caption{Ransom demand distribution for Medusa (Q1 2025).}
\label{fig:medusa_profile}
\end{figure}

To illustrate the granularity of \model{}, we present a focused profile of the Medusa ransomware syndicate in Q1~2025. Unlike groups that conceal negotiations, Medusa adopts a ``public auction'' model, openly publishing ransom demands to intensify extortion pressure.

\textbf{Financial Business Model.}
Medusa exhibits a median ransom demand of \$320{,}000, reflecting a deliberate focus on middle-market victims with sufficient liquidity but limited operational resilience. The wide gap between the median and the \$4.0M maximum indicates a tiered pricing strategy, ranging from deadline extensions to full data deletion guarantees.

\textbf{Sectoral Victimology.}
Cross-referencing ransom demands with \texttt{SECTOR} entities shows a concentration on disruption-sensitive industries, led by Healthcare and Construction, followed by public-sector and educational organizations. This targeting aligns with Medusa’s pricing logic, where demands are calibrated to be financially coercive yet realistically payable under prolonged downtime.

% --- SECTION 6: CONCLUSION ---
\section{Conclusion and Future Work}
\label{sec:conclusion}

In this work, we presented \model{}, an open-source resource for extracting strategic
Cyber Threat Intelligence (CTI) from high-velocity social media streams. Unlike prior
social media CTI corpora, which annotate relevance or technical indicators, \model{}
targets strategic pivots—Target, Sector, Actor, and Impact—that are directly relevant
to automated risk assessment and decision support.

Through comprehensive benchmarking, we show that domain- and medium-adapted encoder models, particularly DarkBERT (89.33\% Strict F1), outperform both general-purpose encoders and large generative models, achieving an approximately 15-point Strict F1 advantage while operating at orders-of-magnitude lower latency. Using a European case study covering H1 2025, we further demonstrate the practical value of \model{} by reconstructing regional threat dynamics and identifying the escalation of the SafePay ransomware campaign well in advance of its consolidation in official threat landscape reports.

\subsection{Limitations and Ethical Considerations}
Our study acknowledges two primary limitations. First, the current dataset is restricted to English-language content, potentially under-representing threats discussed exclusively in Russian or Chinese cybercriminal communities. Second, reliance on a single platform (X/Twitter) introduces a visibility bias towards "loud" extortion campaigns while potentially missing stealthier operations discussed on closed forums.

Ethically, we collected data only from publicly accessible posts using Apify, restricting collection to public endpoints and excluding any private content. To support reproducibility while respecting platform redistribution constraints, we release tweet IDs, span annotations, and code; we do not redistribute tweet text.

\subsection{Strategic Utility}
Beyond the analyses presented in this paper, the \model{} dataset enables a broad class of high-impact strategic intelligence applications. The structured entities support complex attribution and risk analysis tasks, such as identifying which industry sectors are systematically targeted by specific ransomware cartels or estimating the average ransom demand per sector. In addition, the dataset constitutes a high-quality instruction-tuning resource for grounding Large Language Models in real-world Cyber Threat Intelligence, substantially reducing domain-specific hallucinations in automated analytical and report-generation workflows.

\subsection{Future Directions}
Future work will expand the \model{} framework along two complementary axes. First, we aim to construct a Temporal Knowledge Graph~\cite{li2021attackg} by extracting relationships between entities (e.g., \texttt{ACTOR} $\rightarrow$ \texttt{TARGET}), enabling predictive analysis of campaign trajectories. Second, we plan to integrate cross-platform sources, such as dark-web leak sites and Telegram channels, to quantify the ``leak lag''—the time delta between a private extortion demand and its public disclosure. By bridging the gap between unstructured social chatter and structured intelligence, \model{} lays the foundation for the next generation of proactive, automated cyber defense.

% --- APPENDIX ---
\appendix
\section{Evaluation Protocol}
\label{app:evaluation_protocol}

All models are evaluated under a unified span-based protocol to ensure comparability across discriminative and generative architectures.

For encoder-based models, entity spans are produced directly by token-level decoding. For generative models, extraction is performed via instruction prompting with constrained JSON outputs. Predicted entities are retained for scoring only if their surface forms correspond exactly to substrings of the input tweet; non-alignable outputs are discarded.

Inference latency is measured under steady-state batched inference on a single NVIDIA A100 GPU with CUDA synchronization enabled. Reported values represent amortized milliseconds per sample averaged over multiple runs after warm-up, reflecting throughput-oriented deployment scenarios rather than interactive single-sample latency.

\section{LLM Hyperparameters and Prompts}
\label{app:llm_details}
\subsection{System Prompt for Zero-Shot Inference}

\begin{figure}[H]
\centering
\fbox{
\begin{minipage}{0.95\columnwidth}
\small
\raggedright
\texttt{You are an expert Cyber Threat Intelligence (CTI) Automated Analyst. Your task is to extract Named Entities from security alerts (Tweets) and return them in a strict JSON format.}

\vspace{0.5em}
\textbf{\#\#\# ENTITY DEFINITIONS:}
\begin{itemize}
    \setlength\itemsep{0em}
    \item \texttt{TARGET: Victim organization (e.g., "Boeing").}
    \item \texttt{ACTOR: Threat actors (e.g., "LockBit").}
    \item \texttt{SECTOR: Industry (e.g., "Finance").}
    \item \texttt{LOCATION: Countries/flags (e.g., "USA", [Flag Emoji]).}
    \item \texttt{SIZE: Data volume (e.g., "50GB").}
    \item \texttt{DATA\_TYPE: Compromised assets (e.g., "passwords").}
    \item \texttt{PRICE: Ransom/sales (e.g., "\$50,000").}
    \item \texttt{DATE: Specific dates only. IGNORE relative terms.}
\end{itemize}

\textbf{\#\#\# CRITICAL CONSTRAINTS:}
\begin{enumerate}
    \setlength\itemsep{0em}
    \item \texttt{OUTPUT FORMAT: Return ONLY a raw JSON object.}
    \item \texttt{EXACT MATCHING: Match input string exactly.}
    \item \texttt{HASHTAGS: Extract text WITHOUT the hash symbol.}
    \item \texttt{NO ENTITIES: If none found, return \{"entities": []\}.}
\end{enumerate}
\end{minipage}
}
\caption{System Prompt used for Zero-Shot LLM Benchmarking.}
\label{fig:system_prompt}
\end{figure}

\subsection{Fine-Tuning Configuration}
For Generative Large Language Models (Llama-3, Qwen-2.5, Gemma-2), we utilized QLoRA to fine-tune on a single A100 GPU. Table~\ref{tab:qlora_params} details the specific hyperparameters used to ensure stability and prevent overfitting on the small dataset.

\begin{table}[H]
\centering
\caption{Hyperparameters for QLoRA Fine-Tuning.}
\label{tab:qlora_params}
\begin{tabular}{ll}
\toprule
\textbf{Parameter} & \textbf{Value} \\ \midrule
Quantization & 4-bit NormalFloat (NF4) \\
LoRA Rank ($r$) & 16 \\
LoRA Alpha ($\alpha$) & 32 \\
LoRA Dropout & 0.05 \\
Target Modules & \texttt{q\_proj, k\_proj, v\_proj, o\_proj} \\
Learning Rate & $2 \times 10^{-4}$ \\
LR Scheduler & Cosine \\
Batch Size & 1 (with Gradient Accumulation = 16) \\
Max Sequence Length & 1024 \\
\bottomrule
\end{tabular}
\end{table}

%
% ---- Bibliography ----
%

\bibliographystyle{unsrt}
\bibliography{references}

\end{document}